\documentstyle[11pt,emulateapj,epsfig,graphicx]{article}
\def\etal{{\sl~et\,al.~}}

\begin{document} 

\title {A Spectroscopic Redshift for the Cl0024+16 Multiple Arc System:
Implications for the Central Mass Distribution} 

\author{Tom Broadhurst, Xiaosheng Huang, Brenda Frye, Richard Ellis}
\affil{Department of Astronomy, University of California, Berkeley, CA  94720}
\affil{Institute of Astronomy, Madingley Road, Cambridge, CB3 0HA, UK}

%\vskip 24pt
%\clearpage
%\centerline
%{\bf Abstract}

\begin{abstract}

  We present a spectroscopic redshift of $z=1.675$ for the well-known
multiply lensed system of arcs seen in the $z=0.39$ cluster Cl0024+16.
In contrast to earlier work, we find that the lensed images are
accurately reproduced by a projected mass distribution which traces
the locations of the brightest cluster ellipticals, suggesting that
the most significant minima of the cluster potential are not fully
erased.  The averaged mass profile is shallow and consistent with
predictions of recent numerical simulations.  The source redshift
enables us to determine an enclosed cluster mass of
M($<$100kpc/h)=$1.11\pm0.03\times 10^{14}h^{-1}$M$_{\odot}$ and a
mass-to-light ratio of M/L$_B$($<$100kpc/h)=320h(M/L$_B$)$_{\odot}$,
after correction for passive stellar evolution. The arc spectrum
contains many ionized absorption lines and closely resembles that of
the local Wolf-Rayet galaxy NGC4217. Our lens model predicts a high
magnification ($\simeq$20) for each image and identifies a new pair of
multiple images at a predicted redshift of z=1.3.

\end{abstract}

\keywords{cosmology: gravitational lensing --- cosmology: observations ---
galaxies: clusters: individual (Cl0024+16) --- galaxies: distances and
redshifts}

%\clearpage
\section
{Introduction}

  The $z=0.39$ rich cluster Cl0024+16 (Zwicky 1959) displays one of
the finest examples of gravitational lensing. Four clearly related
images are identified around the tangential critical curve in HST
WF/PC-1 data (e.g. Smail \etal 1996). A further radially directed
image of the same source was later found in a refurbished HST WFPC-2
image by Colley \etal (1996). These arcs have been used by Colley
\etal (1996) to construct an image of the source and by Tyson \etal
(1998) to examine details of the mass distribution.

  For many years the redshift of this lensed source has eluded
identification despite long exposures on large telescopes.  The blue
colour and lack of optical emission lines suggests a redshift
$1<$z$<$2 (Mellier \etal 1991). The importance of the redshift for
lensing studies lies primarily in measuring the central mass and
mass-to-light ratio of the lensing cluster, and in the case of
Cl0024+16 these quantities can be measured particularly accurately for
this cluster because of the symmetric arrangement of the images. The
results will depend on the source redshift through its effect on the
ratio of lens to source separations, $d_{LS}/d_S$.  For a nearby lens
this ratio saturates very quickly with increasing source redshift, but
for a more distant lens like Cl0024+16 there is a larger range of
$d_{LS}/d_S$ and hence a larger uncertainty in the
mass, depending on the source redshift.

\section {Observations}
 
  The HST imaging data used here is that obtained by Colley \etal
(1996). The archival data was first reduced for the purposes of
selecting lensed targets for mulislit spectroscopy. Briefly, the
images comprise seven exposures totaling 8400s in the F450W band and
six exposures totaling 6600s in the F814W band. 
Images were aligned to the nearest integer
pixel and cosmic-ray rejected to obtain an average flux. With
multislits on LRIS at the 10m Keck II telescope a 15\arcsec slit was
centered on a high surface-brightness feature of the largest, and
hence most magnified arc (upper end of arc A, see Fig~2a) to maximize
the detection of spectral features. The total exposure time was 80
minutes in 0.8 arcsec seeing, using the 300 line grating blazed at
5000\AA, providing a useful wavelength range of 4500\AA-9500\AA. The
resulting spectrum is shown in Fig~1, revealing many absorption lines,
with a redshift of z=1.675.  The spectrum, observed at an average
resolution of 6\AA\  in the restframe matches closely that of the nearby
starburst galaxy NGC4214 (Leitherer \etal 1996), showing a similar
continuum slope and common absorption lines of SiII 1527\AA, CIV
1550\AA, FeII 1608\AA,2344\AA,2382\AA, 2600\AA, AlII 1671\AA, AlIII
1859\AA. We also detect a foreground $z$=0.18 blue dwarf galaxy
(Fig~2a) which appeared initially to be related to the lensed arcs but
is difficult to reproduce in the lens model (Fig~2b).
 
\section{Modelling the Cluster Lens}

  Lens models for Cl0024+16 have generally improved with higher
quality imaging. Using ground based data, Kassiola \etal (1994) and
Wallington \etal (1995) reproduced fits to the close triplet of arcs
(A,B,C of Figure 2) but considered arc D an unlikely counter
image. Subsequently HST images revealed that A,B,C and D are
morphologically similar in detail (Smail \etal 1996) and that a
further radially directed arc in the cluster center, E, 
is another complete image of the same
source (Colley \etal 1996). Recently a 512 parameter fit to the
resolved imaging data for the arc system has been presented by Tyson
\etal (1998). This solution required the inclusion of a number of
small dark deflecting `mascons' around each of the images to offset
the symmetry of a dominant central potential (see Fig~2 of Tyson \etal
1998).

  To investigate the uncertainties in mass we revisit the mass model
using a simple approach which we nonetheless find sufficient to
reproduce the basic properties of the image configuration. We start by
assigning profiles to the brightest cluster members using the form
advocated by NFW (Navarro, Frenk and White 1995), which has a
characteristic scale, $r_s$, and dimensionless normalization relative
to the cosmological critical density, $\delta_c=\rho_s/\rho_{crit}$,
and allows a wide range of mass concentrations.
Integrating the mass along a column, z, where
$r^2=({\xi_r}{r_s})^2+z^2$ gives
$$\rm{M}(\xi_r)={\rho_s}{r_s^3}(\xi_r)\int_o^{\xi_{r}}d^2\xi_r
\int_{-\infty}^{\infty}
{\frac{1}{(r/r_s)(1+r/r_s)^2}}{dz\over{r_s}}$$ resulting in a
deflection angle $\alpha(\theta)=\frac{4GM(<\theta)}{c^2\theta{d_L}}
\frac{d_{LS}}{d_S}$ in the image plane at position
$\vec{\theta}=\vec\xi_r{r_s}/d_L$.

  Only the brightest 8 cluster members need be included to generate an
accurate fit (all cD galaxies, see Fig~2a) with $r_s$ and $\delta_c$
unconstrained, corresponding to a simple vector addition of deflection
fields. Inclusion of many fainter members produces noise as the fit
rapidly becomes unconstrained for the small number of independent
constraints (5 images).  The fit is achieved with the ``downhill
simplex'' algorithm (Press \etal) by minimizing the difference between the
model predicted locations of the three obvious features (HII regions)
common to the 5 main images of the source
i.e. $$\chi^2=\sum_{k}\sum_{i,j,i>j}\left(({\vec\theta_{i,k}}-{\vec\alpha{(\vec\theta_{i,k})}})
-({\vec\theta_{j,k}}-{\vec\alpha{(\vec\theta_{j,k})}})\right)^2$$ a sum over
all k points of all images.

  The projected (2-D) mass density contours shown in fig~2 are 
modulated by the cluster members despite the projection. The fit
favours an overall shallow profile centered on the central tight clump
of luminous galaxies (fig~2a,b). The two outer ellipticals are seen to be
responsible for the largest departure from symmetry generating the
triple images A,B,\& C. The solution although good is not of course
unique, since a set of discrete profiles generates a fairly 
smooth potential (see Fig~2b) and for this reason we do not
need to introduce additional unknown parameters to describe a diffuse
component. We can convert the the azimuthally-averaged projected model
slope of $\theta^{-0.55}$ in the vicinity of 100kpc/h for comparison
with an NFW profile. The conversion to a real space slope is
$\gamma=-1-2\frac{\xi_r}{1+\xi_r}$, or $\gamma=-1.26$ at the critical
radius. This is very close to the NFW expectation for massive clusters
which have a predicted slope of $\sim-1.3$ at $\sim$100kpc/h (NFW,
Ghingna \etal 1998) corresponding to a combination of
$r_s\sim400$kpc/h and $\delta_c\sim8000$.

\section{Mass/Light Ratios}

  The rest-frame luminosities of the cluster galaxies are converted
from data numbers to an ST-magnitude using header information. The
instrumental colours of the bright ellipticals are very similar with a
mean of $V_{450W}-I_{814W}$=1.62 (2.818 in the AB system or 3.334
normalised to Vega) which in the Johnson system $B-I$=3.7 (Holzman
\etal 1996) and corresponds well to a passively evolved old stellar
population which at the observed redshift $z=0.39$, with $z_f=3$ and
$\tau=.01$ (Bruzual \& Charlot 1995) predicts colours of $B-I$=3.825
(3.727) for $\Omega=0.1$($\Omega=1$) and h=0.5 (solar
metallicity). Correction to the present requires removal of
$0.^m49$--$0.^m52$ of passive evolution.

  The arcs define a convenient radial aperture for comparing mass and
light with a radius of 30.5 arcsec radius intersecting the four bright
tangential images. Using the observed lens and source redshifts, we
normalise the mass distribution by the ratio $d_{LS}/d_S=0.61$
(virtually independent of curvature). Integrating over our model mass
distribution yields $M(\theta)=1.28\times10^{14}h^{-1}M_{\odot}$ for
$\Omega=0.1$, corresponding to a metric radius of $\sim106$kpc/h at the
lens. Note $\Omega$ enters only via the angular-diameter distance of
the lens, so that for $\Omega=1$ the mass is lower by 9\% for the same
angular aperture. The model mass is close to the Einstein mass of a
symmetric lens, $M(\theta<30.5'')_{crit}=1.33\times10^{14}M_{\odot}$,
($\Omega=0.1$ and independent of profile) as expected given the near
circular arrangement of the images about the cluster which means we
have confidence in the mass to an accuracy of less than $\sim$2\%.

  The integrated I-band light in this aperture is $I^{ST}_{814W}=16.2$
or $L_B(\theta)=3.94\times10^{11}h^{-2}L_{\odot}$ ($\Omega=0.1$) after
subtracting passive evolution above and hence, the central a
mass-light ratio is, $M/L_B(\theta)=324h(M/L_B)_{\odot}$ at $z$=0
(using $M_{B\odot}=5.48$). This is slightly higher than other lensing
clusters (Kneib \etal 1996, Natarajan \etal 1998), but note that
neglecting evolution reduces $M/L_B$ by 50\%, and should be allowed
for in accurate comparisons between clusters.  This value is short of
that necessary for closure estimated from local redshift surveys in
the same passband estimated 
requires $(M/L_B)_{crit}=1500^{+700}_{-400}h(M/L)_{B\odot}$ 
(Efstathiou, Ellis \& Peterson 1988) corresponding to
a larger volume and a later mean galaxy type.

  The true size of the source galaxy after subtraction of the
deflection field of image radius, is found to be $r\sim$0.25\arcsec,
$\sim 20$ times fainter than the tangential images , with an unlensed
luminosity $M_B=-20.85\pm0.7-5\log{h}$ and an apparent magnitude of
$I\sim24.8$, typical of what may be expected for a galaxy at the
measured redshift (see Figure 18 of Bouwens, Broadhurst \& Silk 1998).

\section {New Multiple Images}

  New multiply lensed images may be sought with our lens model to
check and improve upon the model. The unknown redshift
introduces an additional free parameter from the distance dependence in
the bend angle.  Hence to search for new multiple images we need only
take the deflection field $\vec \alpha(\vec \theta)$ defined for the
five arc system at $z_1=1.67$ with $(d_{LS}/{d_S})_{z_1}$ and multiply
by a scalar, $f={(d_{LS}/d_S)_{z_2}}/{(d_{LS}/d_{S})}_{z_1}$,
mimicking the effect of a change in the source redshift.

  In practice it is difficult to securely identify new images, mainly
because galaxies are similarly blue and numerous so that a unique
identification based on only 2 passbands is difficult. Furthermore
arcs which are obviously radial or tangential lie close or straddle
the critical curves and therefore much more magnified than their
counterimages which consequently may be too faint to
detect. Furthermore, for a given mass distribution a source must lie
above some minimum redshift to generate multiple images, corresponding
to z$>$1.0 on average for sources within the einstein ring of
Cl0024+16. Here we claim a secure identification of one new pair of
arcs as shown on figure~2. The relative deflection is 90\% of that of
the main arcs corresponding to a predicted redshift of
z=1.31,z=1.34,z=1.33 for
$\Omega$=0.05,$\Omega$=1,$\Omega+\Lambda$=$0.3+0.7$. In principle then
a sufficiently accurate lens model can produce a geometric constraint
on the cosmological curvature, however many more multiple images of
higher redshift sources need to be identified to make this practical.

\section {Discussion and Conclusions}

  It is clear from the above modeling that some degree of mass
substructure is required in Cl0024+16, contrary to the conclusions of
Tyson \etal (1998).  The large mass-to-light ratio we assign to the
location of the luminous ellipticals is far in excess of isolated
elliptical galaxies, meaning these galaxies simply trace well local
minima of a general potential.  This finding is consistent with the
substructure apparent in all carefully studied lensing clusters,
notably, A2218, A370, AC114, A2390, Cl0939+47 (Kneib \etal 1996, Smail
\etal 1996,  Abdelsalam \etal 1998, Pierre \etal 1996, Natarajan \etal 1997,
Seitz \etal 1996). It is not clear if this level of substructure is in
excess of N-body predictions which, as Ghigna \etal (1997) point out
are certainly underestimates within the central 50Kpc/h where the
problem of ``overmerging'' is still significant despite their superior
dynamic range.

  Our azimuthally averaged mass profile is shallow, in 
good agreement with recent high-resolution CDM N-body simulations,
corresponding to an NFW profile with $r_s\sim$400kpc/h which is four
times greater than the observed einstein ring radius and leads to
large central image magnifications.  Although our model does not
explicitly incorporate a separate diffuse component it can be seen in
Fig~2, that the sum of the eight profiles forms a generally smooth
mass distribution illustrating the degeneracy of this sort of modeling
and hence the redundancy of a separate provision for diffuse matter.

  We have successfully used our lens model to find new multiply lensed
galaxies. It is clear that with more color information further systems
will be distinguished in an iterative procedure where the mass model
is successively refined with the incorporation of the new
images. Relative distance predictions can be made this way for
comparison with measured redshifts allowing, in principle, a geometric
measure of the cosmological curvature.

\acknowledgments

We thank Rychard Bouwens and Ben Moore for useful conversations. 
TJB acknowledges NASA grant AR07522.01-96A.

\fontsize{10}{14pt}\selectfont

\begin{figure}[t]
\epsscale{1}
\plotone{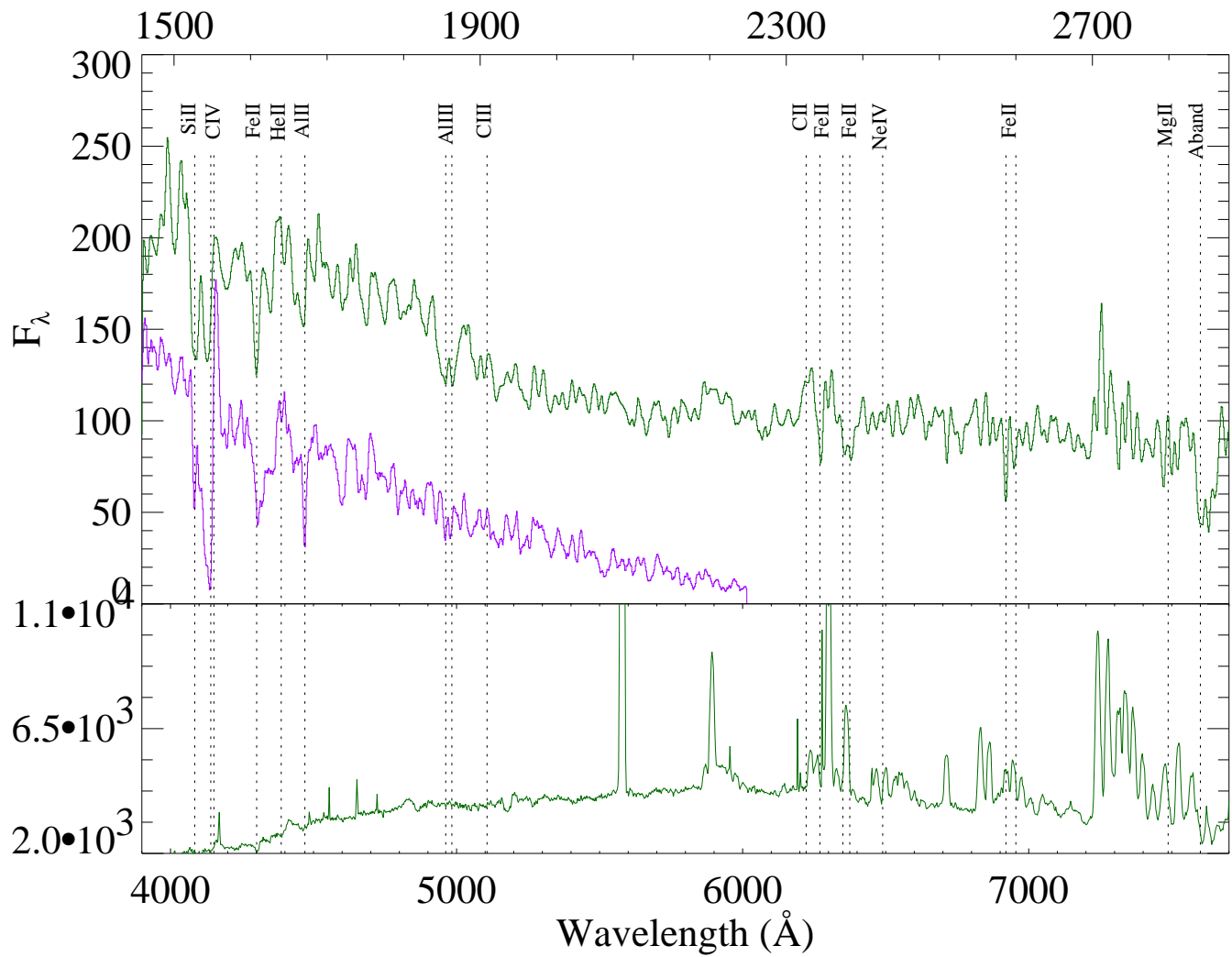}
\caption{The upper curve is fluxed 
spectrum ofi the upper HII region of arc C (fig 2b). Many weak absorption
lines are visible yielding an unambiguous
redshift of z=1.675. Note the similarity with the local ``Wolf-Rayet''
galaxy (lower curve) NGC4214 (Leitherer 1996). The sky spectrum is also
shown, in the lower panel}
\end{figure}

\begin{figure}[t]
\epsscale{1}
\includegraphics[bb=1in 5in 9in 6.in,scale=1.1]{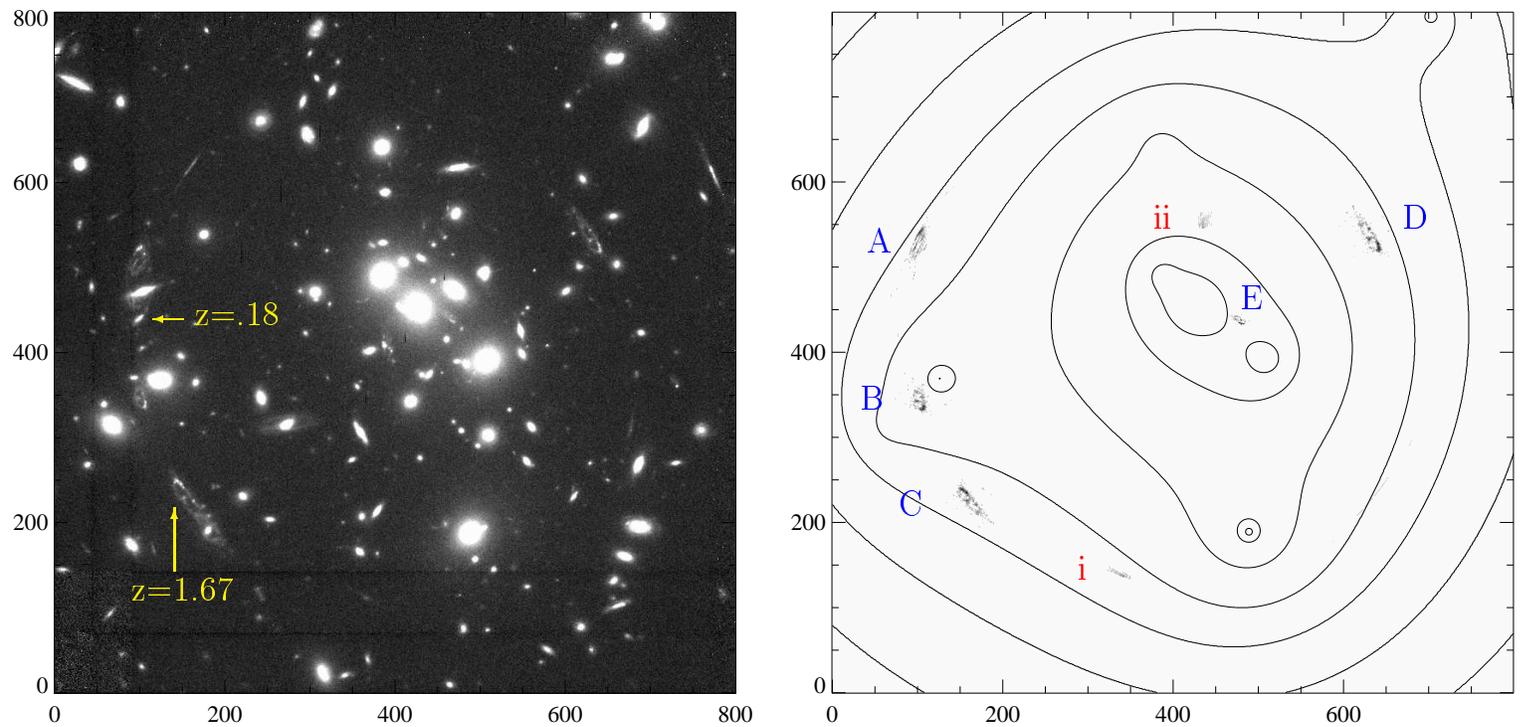}
\caption{ The central region of the
summed B+I band images is shown on the left for comparison with the model
on the right. Note the good agreement with the 5 main images (A-E).
These may be improved further by hand, incorporating masses local
to each image, but will not alter the conclusion that the central mass 
distribution traces well the location of the brightest ellipticals.
A new multiple image pair is identified by the model (i,ii) 
at a predicted redshift of $z\simeq1.3$. The contours represent
surface mass density from $(0.6-1.2)\Sigma_{crit}$, separated by 0.1.}
\end{figure}

\end{document}